\documentclass[conference]{IEEEtran}
\IEEEoverridecommandlockouts
\usepackage{cite}
\usepackage{amsmath,amssymb,amsfonts}
\usepackage{algorithmic}
\usepackage{algorithm}
\usepackage{mathtools}
\usepackage{graphicx}
\usepackage{textcomp}
\usepackage{xcolor}
\usepackage{makecell}
\usepackage{subcaption}

\def\BibTeX{{\rm B\kern-.05em{\sc i\kern-.025em b}\kern-.08em
    T\kern-.1667em\lower.7ex\hbox{E}\kern-.125emX}}

\newcommand{\lFig}[1]{\label{fig:#1}}
\newcommand{\rFig}[1]{Fig. \ref{fig:#1}}
\newcommand{\lSec}[1]{\label{sec:#1}}
\newcommand{\rSec}[1]{Section \ref{sec:#1}}
\newcommand{\lTable}[1]{\label{tab:#1}}
\newcommand{\rTable}[1]{Table \ref{tab:#1}}

\usepackage{eso-pic}
\newcommand\AtPageUpperMycenter[1]{\AtPageUpperLeft{%
 \put(\LenToUnit{0.15\paperwidth},\LenToUnit{-1cm}){%
     \parbox{0.8\textwidth}{\raggedleft\fontsize{9}{11}\selectfont #1}}%
 }}%
\newcommand{\conf}[1]{%
\AddToShipoutPictureBG*{%
\AtPageUpperMycenter{#1}
}
}

\begin{document}

\title{ECO: Edge-Cloud Optimization of 5G applications\\
{\footnotesize \textsuperscript{}}
\thanks{978-1-7281-9586-5/21/\$31.00 $\copyright$2021 IEEE\\DOI 10.1109/CCGrid51090.2021.00078}
}

\author{\IEEEauthorblockN{Kunal Rao}
\IEEEauthorblockA{
\textit{NEC Laboratories America}\\
Princeton, NJ \\
kunal@nec-labs.com}
\and
\IEEEauthorblockN{Giuseppe Coviello}
\IEEEauthorblockA{
\textit{NEC Laboratories America}\\
Princeton, NJ\\
giuseppe.coviello@nec-labs.com}
\and
\IEEEauthorblockN{Wang-Pin Hsiung}
\IEEEauthorblockA{
\textit{NEC Laboratories America}\\
San Jose, CA\\
whsiung@nec-labs.com}
\and
\IEEEauthorblockN{Srimat Chakradhar}
\IEEEauthorblockA{
\textit{NEC Laboratories America}\\
Princeton, NJ \\
chak@nec-labs.com}
}

\maketitle
\conf{2021 IEEE/ACM 21st International Symposium on Cluster, Cloud and Internet Computing (CCGrid)}

\begin{abstract}
Centralized cloud computing with 100+ milliseconds network latencies cannot meet the tens of milliseconds to sub-millisecond response times required for emerging 5G applications like autonomous driving, smart manufacturing, tactile internet, and augmented or virtual reality. 
We describe a new, dynamic runtime that enables such applications to make effective use of a 5G network, computing at the edge of this network,  and resources in the centralized cloud, at all times.
Our runtime continuously  monitors the interaction among the microservices, estimates the data produced and exchanged among the microservices, and uses a novel graph min-cut algorithm to dynamically map the microservices to the edge or the cloud to satisfy application-specific response times. Our runtime also handles temporary network partitions, and maintains data consistency across the distributed fabric by using microservice proxies to reduce WAN bandwidth by an order of magnitude, all in an {\it application-specific manner} by leveraging knowledge about the application's functions, latency-critical pipelines and intermediate data. We illustrate the use of our 
runtime by successfully mapping two complex,  representative real-world video analytics applications to the AWS/Verizon Wavelength edge-cloud architecture, and improving application response times by 2x when compared with a static edge-cloud implementation. 
\end{abstract}


\section{Introduction}

Cloud services are everywhere. From individual users watching over-the-top video content to enterprises deploying software-as-a-service, cloud services are increasingly how the world consumes content and data. Although centralized cloud computing is ubiquitous, and economically efficient, an exponential growth in internet-connected machines and devices
is resulting in emerging new applications, services, and workloads for which the centralized cloud quickly becomes computationally inefficient~\cite{inefficient-cloud}. New, emerging applications like autonomous driving, smart manufacturing, tactile internet, remote surgeries, real-time closed-loop control as in Industry 4.0, augmented or virtual reality require tens of milliseconds to sub-millisecond response times \cite{tactile-interet}. For these applications,  processing all data in the cloud and returning the results to the end user is not an option because it takes too long, uses excessive power, creates privacy and security vulnerabilities, and causes scalability problems. 

The new applications demand a different kind of computing fabric, one that is distributed and built 
to support low-latency and high-bandwidth service delivery capability, which centralized cloud implementations with 100+ milliseconds (ms) network latencies are not well-suited for \cite{cloudization}. For such applications, consideration of a few human-related benchmarks is useful. The time to blink an eye is about 150 ms \cite{eye-blink}, and a world-class sprinter can react within 120 ms of a starting gun \cite{sprinter}. The human auditory reaction time is about 100 ms, and the typical human visual reaction time is in the range of 10 ms \cite{human-reaction-time}. Also, nerve impulses travel at over 100 meters per second in the human body \cite{nerve-conduction}, and the time required to propagate a signal from hand to brain is about 10 ms \cite{brain-to-arm}. If application response times are in this range, then it is possible to interact with distant objects with no perceived difference to interactions with a local object. 
Even faster, sub-milliseconds response times are required for machine to machine communication as in industry 4.0, where closed-loop real-time control systems automate processes like quality control, high-speed manufacturing, food packaging, construction, and e-health, without the involvement of any human.  Clearly, if applications are to approach this level of responsiveness, the network latencies (not including processing) must be much less than 100 ms, which 5G network and edge computing can deliver.

\subsection{Edge-cloud}
The speed of light places a fundamental limit on the network latencies, and  the farther the distance between the data source and the processing destination, the more time it will take to transmit the data to the destination. So, edge computing places computing resources at the edges of the Internet in close proximity to devices, information sources  and end-users, where the content is created and consumed. This, much like a cache on a CPU, increases bandwidth and reduces latency between the end-users or data sources, and data processing. Today, centralized cloud has more than 10,000 data centers \cite{data-center} scattered across the globe, but within the next five years, driven by a need to get data and applications closer to end-users (both humans and machines), orders of magnitude more heavily scaled-down data centers are expected to sprout up at the edge of the Internet to form the {\it edge-cloud}.
 A tiered system, as shown in ~\rFig{edge-cloud}, with the cloud, and additional, heterogeneous computing and storage resources placed inside or in close proximity to the sensor, is emerging as a computing reference architecture for edge-cloud applications. 
 The cloud is expected to be used for a fully centralized application delivery, management and execution of select application functions that may require a global perspective. 
 \begin{figure}[t]
 \centering
    \includegraphics[width=0.95\linewidth]{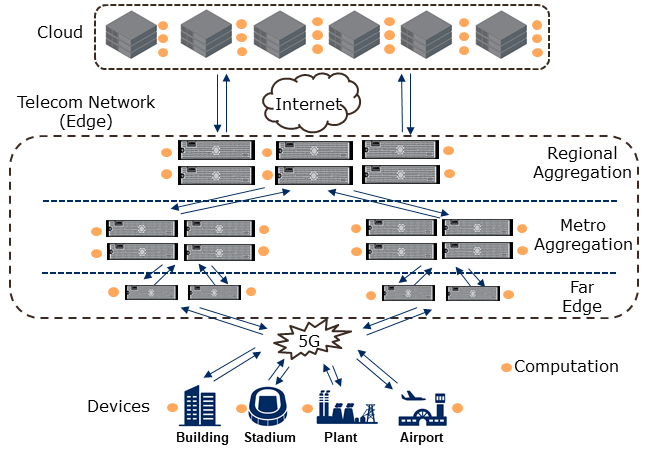}
    \caption{Tiered, edge-cloud reference architecture}
    \lFig{edge-cloud}
\end{figure}

 
 
\subsubsection{Advantages}
A tiered reference architecture is attractive for several reasons. 
\textit{First}, 
computing capability at the edge of the cellular network can enable fundamentally new applications that require high data-rate instantaneous communications, low latency, and massive connectivity, which can be provided by new networks like 5G. 
\textit{Second}, by extending the cloud paradigm to the heavily scaled-down edge data centers, it is possible for edge providers to quickly develop, install, deliver and manage applications using the same tools and techniques that are used in the cloud.
\textit{Third}, cloud can be used for a fully centralized application delivery, and management, in addition to providing computing resources for execution of application functions that require a global perspective.

 

\subsubsection{Challenges}
Despite its promise, and obvious advantages, the tiered reference architecture also poses several fundamental challenges for applications:
\textit{First}, 
mapping and execution of applications to continuously meet low-latency response times on a dynamic, geo-spatially distributed, and complex edge-cloud infrastructure with heterogeneous resources (different types of networks and computing resources) is a major challenge. 
\textit{Second}, edge resources (compute, storage and network bandwidth) being severely limited, shared across applications and more expensive than cloud resources, need special strategies
to realize economically viable low-latency response applications. \textit{Third}, handling temporary network disruptions in a distributed infrastructure to meet low-latency application response need techniques beyond traditional application-agnostic methods. 

\subsubsection{Our contribution}
In this paper, we 
describe a runtime that enables applications to make effective use of the large-scale distributed platform consisting of a 5G network, computing and storage resources across the cloud, different tiers of edge-cloud, and the devices. 
Our runtime leverages internal knowledge about the application's  microservices, their interconnections, and the critical pipelines of microservices that determine the latency response of the application,
to dynamically map the microservices to different tiers of computing and storage resources to achieve application latency goals, maintain data consistency across the distributed storage by using microservice proxies to reduce WAN bandwidth by an order of magnitude, and finally, handle temporary network disconnections, all in an {\it application-specific manner}. Our runtime continuously  monitors data produced and exchanged among the microservices, and uses a novel graph min-cut algorithm to efficiently map the microservices to the edge or the cloud.
We illustrate the use of  the proposed runtime by successfully mapping two different types of real-world video analytics applications to the AWS/Verizon Wavelength edge-cloud architecture (discussed in \rSec{test-bed}). The proposed approach is just as easily applicable to any application that will be packaged in a 5G network slice~\cite{5G-network-slicing}, whose definition is expected to be standardized in the near future.    

\section{Microservices and Critical pipelines}
Microservices \cite{microservices} are increasingly becoming popular, especially in cloud services. They are independently deployable with an automated deployment mechanism, they need a bare minimum of management, they can be built in different programming languages and employ different data storage technologies, and each microservice can be independently updated, replaced and scaled. We focus on such microservices based applications, and as part of the specification of the application, we expect the developer to specify the critical pipelines that determine the response latency of the application from measurement to action, and the acceptable response latencies for the critical pipelines. Applications can have a large number of microservices, with complex interactions, and multiple latency-critical pipelines. Please note that individual latencies of the microservices do not have to be specified, only the desired aggregate latency of the critical pipelines. These microservices are packaged as  docker images, and they run as docker containers inside pods in a third-party orchestration framework like  Kubernetes\cite{kubernetes}. We consider two different types of real-world video analytics applications and show the critical pipelines for each of them.

\begin{figure}[b]
\centering
    \includegraphics[width=0.95\linewidth]{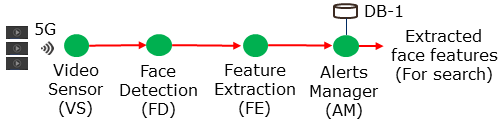}
    \caption{Investigation and forensics application pipeline}
    \lFig{search-pipeline}
\end{figure}

\subsection{Investigation and forensics}
\lSec{investigation-forensics}
This application is required by law enforcement agencies to quickly search past history of suspects or criminals. 
In this application, days or weeks worth of videos need to be processed within few minutes or hours, so that the required information can be made available as soon as possible, to speed up the investigation process. As a first step towards analyzing these videos, they have to be transferred very quickly for the processing to begin. Next, as the video file processing is on-going and intermediate results are ready, they are continuously made available as soon as possible for further investigative actions. This is a bandwidth-sensitive application; the video file transfer depends on the available upload bandwidth and the continuous reception of results depends on the available download bandwidth. \rFig{search-pipeline} shows the various microservices involved in this application and the critical pipeline (shown in red). The video files are read by Video Sensor (VS) microservice and frames from these files are made available for further processing. Note that VS can split files and make frames available in batches as well. These frames are then processed by Face Detection (FD) followed by Feature Extraction (FE) microservices to detect faces and extract unique facial feature templates. These extracted features are then made available for search through Alerts Manager (AM) microservice, which stores and retrieves facial features from a persistent store (DB-1).
All microservices i.e. VS, FD, FE and AM form the critical pipeline, which determine the response latency of the application, from measurement (i.e. receipt of a video file) to action (i.e. when faces are made available as they are extracted). 

\begin{figure}[t]
\centering
    \includegraphics[width=0.95\linewidth]{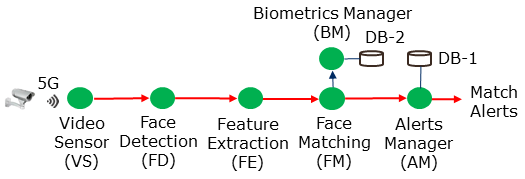}
    \caption{Real-time monitoring application pipeline}
    \lFig{real-time-monitoring-pipeline}
\end{figure}

\begin{figure}[t]
\centering
    \includegraphics[width=.95\linewidth]{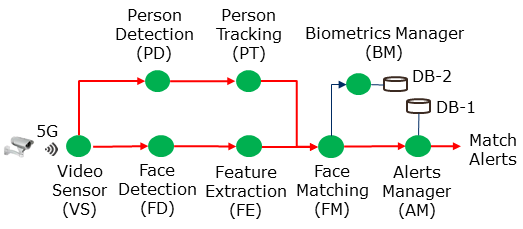}
    \caption{Application pipeline for variant of real-time monitoring}
    \lFig{admitlist-pipeline}
\end{figure}


\subsection{Real-time monitoring}
This is a typical surveillance application where video cameras are used to monitor a particular facility and face-recognition technology is used to flag undesirable individuals like criminals, or to control access and allow only authorized individuals (say, employees) into the facility. 
This is a latency-sensitive application; the alert is generated as soon as the person of interest is seen (possibly within milliseconds). 
\rFig{real-time-monitoring-pipeline} shows the entire application as a collection of interconnected microservices and the critical pipeline.
The Biometrics Manager (BM) microservice manages names and pictures of individuals to be monitored. Video Sensor (VS) microservice receives the video stream, decodes it and makes it available as individual frames for further processing. Faces in these frames are then detected by the Face Detection (FD) microservice. Once faces are detected, unique facial feature template are extracted by the Feature Extraction (FE) microservice. These feature templates are then matched against the pre-registered individuals' facial feature template by Face Matching (FM) microservice. 
Based on the match between these facial templates, an alert is generated which is managed by Alerts Manager (AM) microservice. 
The critical pipeline (shown in red) in this application consists of VS, FD, FE, FM and AM microservices, which determine the response latency of the application, from measurement (i.e. capture of a frame by a video camera) to action (i.e. when an alert is produced). 

A common, but more complex variant of the real-time monitoring application is shown in \rFig{admitlist-pipeline}, which involves Person Detection (PD) and Person Tracking (PT) microservices to track individuals when their faces are not always facing the cameras. This application has 2 critical pipelines (shown in red) (a) VS, PD, PT, FM, AM and (b) VS, FD, FE, FM, AM. 

Note that all the above applications include microservices that maintain state information in databases. For example, microservices AM and BM each maintain global state in separate databases, and interaction with the databases is not in the critical path of the application. Also, note how the complexity of applications progressively grows from \rFig{search-pipeline}, where there is a single critical path and single DB, to \rFig{real-time-monitoring-pipeline}, with single critical path but two DBs and finally \rFig{admitlist-pipeline}, with multiple critical paths and multiple DBs. All these, and even more complex applications are handled by our runtime, which is discussed next in \rSec{runtime}.

\section{Runtime for edge-cloud}
\lSec{runtime}
Our runtime consists of two components i.e. Policy Engine (PE) and Scheduler (S), which are themselves implemented as microservices and run as containers within Kubernetes. 

\subsubsection{Policy Engine (PE)}
\lSec{policy-engine}
This microservice continuously monitors application-level as well as network-level performance metrics, and uses that to determine appropriate partitioning of application in the edge-cloud infrastructure. 
Each application in our case is a set of microservices, which are chained together in some manner to form a topology or a graph $G=(V,E)$, where the set of vertices $V=(v_1,v_2,..v_n)$ denotes the microservices and edge $e(v_i,v_j) \in E$ represents the communication between microservice $v_i$ and $v_j$, where $v_i$ and $v_j$ are neighbors. Each vertex $v\in V$ is assigned with two cost weights $w(v)^{edge}$ and $w(v)^{cloud}$, which are the cost of running the microservice on the edge and cloud respectively. Cost of running microservice $v$ in the edge is given by \eqref{edge-cost} and cost of running it in the cloud is given by \eqref{cloud-cost}. 

\begin{equation}
w(v)^{edge} = T_v^{edge} * P_v^{edge}
\label{edge-cost}
\end{equation}

\begin{equation}
w(v)^{cloud} = T_v^{cloud} * P_v^{cloud}
\label{cloud-cost}
\end{equation}

where $T_v^{edge}$ is the execution time of microservice $v$ on the edge, $P_v^{edge}$ is the price (AWS cost) of running the microservice on the edge, $T_v^{cloud}$ is the execution time of the microservice on the cloud and $P_v^{cloud}$ is the price (AWS cost) of running the microservice in the cloud. Note that some microservices cannot be offloaded to the cloud and they have to remain in the edge e.g. microservices that receive inputs from devices in the carrier network or microservices that deliver data to devices in the carrier network. Such microservices are bound to the edge and they only have edge costs.

Each vertex receives one of the two weights depending on where it is scheduled to run i.e. it will get a weight $w(v)^{edge}$ if it is scheduled to run in the edge or $w(v)^{cloud}$ if it is scheduled to run in the cloud. Each edge $e(v_i,v_j) \in E$ represents the communication between $v_i$ and $v_j$, where $v_i$ is in the edge and $v_j$ is in the cloud (or vice versa), and this edge is assigned a weight given by \eqref{communication-cost}

\begin{equation}
w(e(v_i,v_j)) = \frac{data\_in_{i,j}}{bw_{upload}} + \frac{data\_out_{i,j}}{bw_{download}}
\label{communication-cost}
\end{equation}

where $data\_in_{i,j}$ is the amount of data transferred (uploaded) from $v_i$ to $v_j$, $data\_out_{i,j}$ is the amount of data received (downloaded) from $v_j$ to $v_i$, $bw_{upload}$ is the network upload bandwidth and $bw_{download}$ is the network download bandwidth between edge and cloud. Note that when $v_i$ and $v_j$ are both either in the cloud or in the edge, then communication latency between them is considered zero (given by  \eqref{total-latency} and \eqref{f-values}). 

The total latency for the application is the end-to-end time for processing a unit of work, which depends on the time taken by the microservices in the critical pipeline in the application. This 
total latency is given by \eqref{total-latency}


\begin{equation}
\begin{split}
L_{total} & = {\sum\limits_{v \in V} F_v\times{T_v^{edge}}} 
+ {\sum\limits_{v \in V} (1 - F_v)\times{T_v^{cloud}}} \\
&+ {\sum\limits_{e(v_i,v_j) \in E} F_e\times{w(e(v_i,v_j))}}
\end{split}
\label{total-latency}
\end{equation}

where $L_{total}$ is the sum of the edge latency i.e. processing time taken by microservices on the edge for a unit of work, cloud latency i.e. processing time taken by microservices on the cloud for a unit of work and communication latency i.e. time taken for data transfer between the edge and the cloud. Flags $F_v$ and $F_e$ in \eqref{total-latency} are defined as follows:

{\small
\begin{equation}
    F_v =
    \begin{cases}
        1, &\text{if } v  \in V^{edge} \\
        0, &\text{otherwise}
    \end{cases}
    \text{and }
    F_e =
    \begin{cases}
        1, &\text{if } e  \in E_{cut} \\
        0, &\text{if } e \notin E_{cut}
    \end{cases}
    \label{f-values}
\end{equation}
}

where $V^{edge}$ is the set of vertices (microservices) scheduled to run on the edge and $E_{cut}$ is the set of edges $e(v_i,v_j)$ in which $v_i$ and $v_j$ are scheduled on edge and cloud or vice versa. 

We formulate the total cost given by \eqref{total-cost}


\begin{equation}
\begin{split}
Cost_{total} & = c_{edge}\times{\sum\limits_{v \in V} F_v\times{w(v)^{edge}}} \\ 
&+ c_{cloud}\times{\sum\limits_{v \in V} (1 - F_v)\times{w(v)^{cloud}}}
\end{split}
\label{total-cost}
\end{equation}

where the total cost is the sum of the edge computation cost and cloud computation cost, and weight parameters $c_{edge}$ and $c_{cloud}$ can be used to adjust the relative importance between them. The goal of partitioning is to find a cut in the graph $G=(V,E)$ with minimum total cost under given total latency constraint per unit of work. This latency constraint is part of application specification, and PE adheres to this constraint while determining the cut in the graph. This cut separates the graph into two disjoint sets, where one side of the cut is scheduled on the edge while the other side is scheduled on the cloud.

\begin{figure}[b]
\centering
    \includegraphics[width=0.75\linewidth]{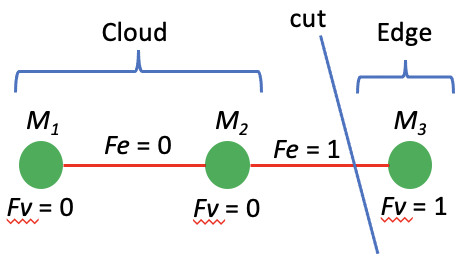}
    \caption{Microservices partitioning example}
    \lFig{cut-example}
\end{figure}

As an example, consider an application with three microservices $M_1$, $M_2$ and $M_3$ chained in a sequential manner as shown in \rFig{cut-example}. All three microservices form part of the critical pipeline. Let execution times of $M_1$, $M_2$ and $M_3$ on edge be ${T_{M_1}}^{edge}$, ${T_{M_2}}^{edge}$ and ${T_{M_3}}^{edge}$ respectively, while execution times on cloud be ${T_{M_1}}^{cloud}$, ${T_{M_2}}^{cloud}$ and ${T_{M_3}}^{cloud}$ respectively. Let the cost of running $M_1$, $M_2$, and $M_3$ on edge be $w({M_1})^{edge}$, $w({M_2})^{edge}$ and $w({M_3})^{edge}$ respectively, while the cost of running in the cloud be $w({M_1})^{cloud}$, $w({M_2})^{cloud}$ and $w({M_3})^{cloud}$ respectively. Also, let the communication latency between $M_1$ and $M_2$ be $w(e({M_1},{M_2}))$, while the communication latency between $M_2$ and $M_3$ be $w(e({M_2},{M_3}))$. Based on the partition scheme, if the cut on the graph formed by these microservices is determined to be between $M_2$ and $M_3$, such that $M_1$ and $M_2$ are scheduled on the cloud and $M_3$ is schedule on the edge, then value of $F_v$ for $M_3$ would be 1, while it will be 0 for $M_1$ and $M_2$ (\eqref{f-values}). Also, value of $F_e$ for $w(e({M_2},{M_3}))$ will be 1, while it will be 0 for $w(e({M_1},{M_2}))$ (again \eqref{f-values}). The total latency given by \eqref{total-latency} will then translate to sum of ${T_{M_3}}^{edge}$ (edge latency), ${T_{M_1}}^{cloud}$, ${T_{M_2}}^{cloud}$ (cloud latency) and finally $w(e({M_2},{M_3}))$ (communication latency) (shown in \eqref{m1-m2-m3-latency}). And, the total cost given by \eqref{total-cost} will translate to sum of $w({M_3})^{edge}$ (edge computation cost), $w({M_1})^{cloud}$ and $w({M_2})^{cloud}$ (cloud computation cost) (shown in \eqref{m1-m2-m3-cost}).

\begin{equation}
\begin{split}
L_{{M_1},{M_2},{M_3}} & = {T_{M_3}}^{edge}
+ {T_{M_1}}^{cloud} + {T_{M_2}}^{cloud} \\
&+ w(e({M_2},{M_3}))
\end{split}
\label{m1-m2-m3-latency}
\end{equation}

\begin{equation}
\begin{split}
C_{{M_1},{M_2},{M_3}} & = w({M_3})^{edge}
+ w({M_1})^{cloud} + w({M_2})^{cloud}
\end{split}
\label{m1-m2-m3-cost}
\end{equation}


In scenarios where there are multiple layers of computing infrastructure available, such that the cost reduces as you go to upper layers at the expense of increased latency, the same method can be applied iteratively across layers to identify the appropriate allocation and scheduling of microservices. For example, consider three computing layers A, B and C, with A being at the top, B in the middle and C at the bottom. Cost of computing is lower as you go up from C to A, while the latency is higher as you go up from C to A. In this scenario, our partitioning scheme will first consider C as the edge and B as the cloud for equations \eqref{edge-cost}, \eqref{cloud-cost} and \eqref{communication-cost}. Once this partition is determined, certain microservices will be scheduled to run on C (edge), while others will be scheduled to run on B (cloud). Then, for the microservices that are scheduled to run on B, and only for these microservices, 
our partitioning scheme will be applied again. This time B is considered as the edge and A as the cloud for equations \eqref{edge-cost}, \eqref{cloud-cost} and \eqref{communication-cost}. The set of microservices will now be split to run between B (edge) and A (cloud). This way, the various microservices will be allocated and scheduled to run on layers A, B and C. This iterative process can be extended to any number of computing layers and an effective partitioning of microservices can be realized across the various computing layers. By going in this bottom up fashion, we ensure that latency constraint is met at each iteration and cost goes down in successive iterations.

PE also transparently introduces proxy microservices as needed within the application pipeline. For example, in real-time monitoring application, AM is in the critical pipeline but it also interacts with a persistent storage (database) which stores alerts globally. Having AM and the storage in the cloud, with a global perspective, is ideal but it affects the latency of the critical pipeline. To alleviate this problem, PE introduces Alerts-Manager at Edge (AM-E) microservice, which is a proxy for alerts-manager microservice in the cloud. AM-E receives alerts from application microservices running locally in the edge and it publishes the alerts over a ZeroMQ channel for other applications to consume. These alerts are maintained in a temporary buffer and synchronized in the background with AM microservice's persistent storage in the cloud. Addition of AM-E proxy does not add any overhead, rather it expedites the critical pipeline since delivery of alerts happens directly from the edge and synchronization with AM happens in the background without affecting the critical pipeline.


\begin{algorithm}[t]
\caption{Application scheduling}
\label{application-partitioning}
\begin{algorithmic}[1]
    \WHILE{true}
        \FOR{$app$ $\in$ $apps$}
            \IF{!isAppScheduled($app$) \textbf{OR}
            \STATE conditionsChanged($app$, $a\_p$, $n\_p$)}
                \STATE \hspace{0.2cm} $partition \leftarrow$ getPartition($app$, $a\_p$, $n\_p$)
                \STATE \hspace{0.2cm} scheduleApp($app$, $partition$)
            \ENDIF
        \ENDFOR
        \STATE sleep($interval$)
    \ENDWHILE
\end{algorithmic}
\end{algorithm}

\subsubsection{Scheduler (S)}
\lSec{scheduler}
Scheduler manages the actual scheduling of application components between edge and cloud. These decisions are made (a) statically i.e. at the start of the application and (b) dynamically i.e. while the application is running. 

Before the execution of the application starts, based on application and network parameters i.e. the current network condition, and apriori knowledge about execution times and communication across various microservices of the application, the partitioning determined by PE is used by S to decide where to schedule the various microservices. This apriori knowledge, used at the start, is obtained by profiling the application and identifying the execution times of microservices and communication patterns between them. 

After the application starts running, scheduler S continuously receives application-level and network-level performance data from PE. This is used to periodically check if the previously selected partition is still good or needs to be adjusted dynamically based on the changing environmental (application-level and/or network-level) conditions. If the application performance and/or network performance degrades, resulting in latency constraint for the application not being met by the current partition, then the conditions are considered to have changed and new partitioning is determined to adjust to the changed conditions. 

Algorithm \ref{application-partitioning} shows the scheduling algorithm used by S to schedule microservices in the application. The {\it getPartition(app, a\_p, n\_p)} function uses the formulation and graph min-cut technique described in \rSec{policy-engine} to obtain the partition and {\it scheduleApp(app, partition)} function uses underlying Kubernetes orchestration framework to schedule and run the application. Note that the individual microservices (and their state) are already pre-loaded at different tiers and the orchestration framework can start or stop running any microservice on any tier,  as soon as the signal is received from S.
\section{Experimental results}
\subsection{Test bed for 5G applications in edge-cloud}
\lSec{test-bed}
\begin{figure*}[]
\centering
   \includegraphics[width=0.95\textwidth]{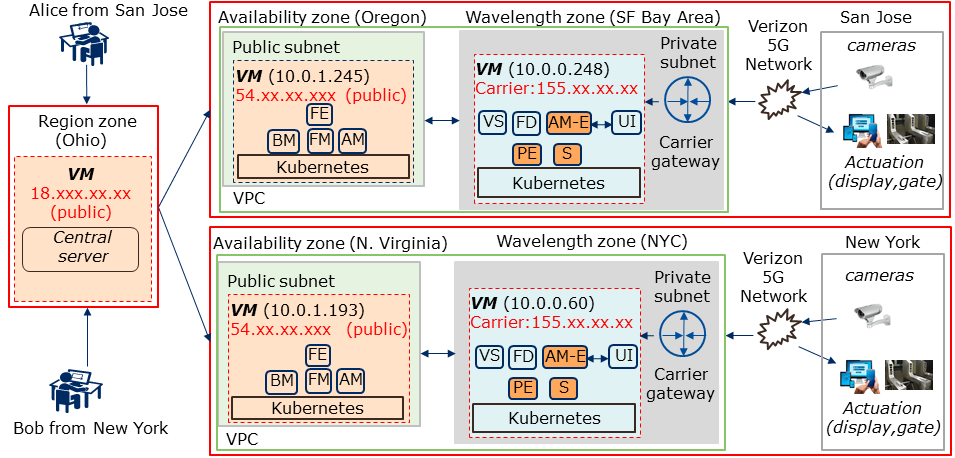}
   \caption{Test bed for 5G applications in edge-cloud}
   \lFig{test-bed}
\end{figure*}
We use AWS Wavelength \cite{aws-wavelength} to explore edge-cloud solutions for 5G applications, in which
AWS compute and storage is placed directly within the communication provider's network. 
This removes the latency that results from multiple hops between regional aggregation sites and across the Internet, and enables new, low-latency applications.
\rTable{aws-wl-instances-types} shows the configuration for various instance types available in AWS Wavelength.
A typical AWS Wavelength infrastructure is shown in \rFig{wavelength-infrastructure}, where the Amazon Virtual Private Cloud (VPC) is extended to include the Wavelength Zone and supported EC2 instances are spawned in the Wavelength zone to handle latency-sensitive  application components. 5G devices in the carrier network connect to the Wavelength Zone through the carrier gateway and network traffic from these devices is directly routed to the VMs in Wavelength zone without leaving the carrier network. 

Our test bed for 5G applications in edge-cloud infrastructure is shown in \rFig{test-bed} and consists of two separate MicroK8s Kubernetes clusters in the Availability (cloud) and Wavelength (edge) Zone. Video feed from cameras is routed into VM in Wavelength zone over Verizon 5G network, and the VMs in Wavelength zones and Availability zones are configured to be in the same VPC. For all our experiments, we used Verizon 5G MIFI 2100 hotspot and t3.xlarge instance type in Amazon AWS infrastructure for Wavelength and Availability Zone.
 
\begin{figure}[t]
\centering
    \includegraphics[width=0.95\linewidth]{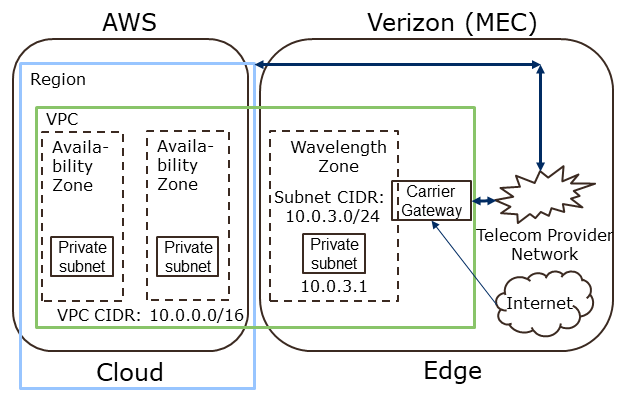}
    \caption{AWS Wavelength Infrastructure}
    \lFig{wavelength-infrastructure}
\end{figure}

\begin{table}[t]
\caption{AWS Wavelength instance types}
\begin{center}
    \begin{tabular}{|l|c|c|c|c|c|c|}
    \hline
     \makecell{Instance \\ Types} & \makecell{vCPUs} & \makecell{Memory \\ (GiB)} & \makecell{GPU} &  \makecell{Network \\ Perf. \\ (Gigabit)} & \makecell{Price \\ (per hour)}\\
    \hline
    t3.medium & 2 & 4 & no & upto 5 & \$0.056\\
    \hline
    t3.xlarge & 4 & 16 & no &  upto 5 & \$0.224\\
    \hline
    r5.2xlarge & 8 & 64 & no &  upto 10 & \$0.68\\
    \hline
    g4dn.2xlarge & 8 & 32 & \makecell{yes} & upto 25 & \$1.317\\
    \hline
    \end{tabular}
\end{center}
\lTable{aws-wl-instances-types}
\end{table}

\subsection{5G and WAN network latencies}
\lSec{network-performance}
In this section we present some raw network performance numbers. Although Verizon claims to have 5G speed in San Francisco bay area, the location where we measured, did not get the full expected 5G speed and therefore we measured at two different locations within the SF Bay area. We report the numbers we observed using this 5G hotspot at these two locations, namely location-1 (GPS coordinates: 37.351368, -121.994782
) and location-2 (GPS coordinates: 37.349002, -121.993945). \rTable{network-performance-location-1} summarizes the latency, upload bandwidth and download bandwidth that we observed between (a) Device in Wavelength zone to VM in Wavelength zone (b) Device in non-wavelength zone (since user can connect from anywhere) to VM in Availability zone and (c) VM in Availability zone to VM in Wavelength zone. We used ping to measure the latency and iperf3 \cite{iperf3} to measure the upload and download bandwidth. We repeat the experiments multiple times and report the minimum, average, maximum and the standard deviation values for all measurements.

\begin{table}[t]
\caption{Network performance at location-1 in SF Bay area}
\begin{center}
    \begin{tabular}{|l|c|c|c|c|}
    \hline
    \makecell{Metrics} & \makecell{Min.} & \makecell{Avg.} & \makecell{Max.} & \makecell{Std. \\ dev.}\\
    \hline
    \text{\makecell{Latency between (ms)}} & & & & \\
    \textit{(a) Device and Wavelength} & 22 & 39.7 & 81 & 9.94\\
    \textit{(b) Device and Availability} & 66.05 & 78.88 & 243.5 & 20.89\\
    \textit{(c) Wavelength and Availability} & 11.4 & 25.8 & 34 & 3.66\\
    \hline
    \text{\makecell{Upload Bandwidth (Mbits/s)}} & & & & \\
    \textit{(a) Device to Wavelength} & 22.2 & 28.43 & 37 & 3.17 \\
    \textit{(b) Device to Availability} & 0.8 & 2.62 & 7 & 1.41\\
    \textit{(c) Wavelength to Availability} & 21 & 35.47 & 71 & 10.09\\
    \hline
    \text{\makecell{Download Bandwidth (Mbits/s)}} & & & & \\
    \textit{(a) From Wavelength to Device} & 59 & 151.3 & 183 & 17.05 \\
    \textit{(b) From Availability to Device} & 9 & 31.74 & 56 & 12.34\\
    \textit{(c) From Availability to Wavelength} & 20.6 & 38.31 & 74 & 10.54\\
    \hline
    \end{tabular}
\lTable{network-performance-location-1}
\end{center}
\end{table}

\begin{table}[t]
\caption{Network performance at location-2 in SF Bay area}
\begin{center}
    \begin{tabular}{|l|c|c|c|c|}
    \hline
    \makecell{Metrics} & \makecell{Min.} & \makecell{Avg.} & \makecell{Max.} & \makecell{Std. dev.}\\
    \hline
    \text{\makecell{Latency (milli-seconds)}} & & & & \\
    \textit{(a) Device to Wavelength} & 24.6 & 30.9 & 50.9 & 4.5\\
    \hline
    \text{\makecell{Upload Bandwidth (Mbits/s)}} & & & & \\
    \textit{(a) Device to Wavelength} & 40 & 42.8 & 46 & 2.4 \\
    \hline
    \text{\makecell{Download Bandwidth (Mbits/s)}} & & & & \\
    \textit{(a) Wavelength to Device} & 255 & 292 & 318 & 18.25 \\
    \hline
    \end{tabular}
\lTable{network-performance-location-2}
\end{center}
\end{table}

From \rTable{network-performance-location-1}, we can see that the latency between device to VM in Wavelength zone is much lower than that between device to VM in Availability zone, indicating faster turn-around time for responses from Wavelength zone compared to those from Availability zone (where the standard deviation is also quite high). Also, the latency between VM in Availability zone and the VM in Wavelength zone is very low, indicating that we can offload processing to the availability zone and get back response from there to Wavelength zone quickly. With respect to upload bandwidth, we see that from device to VM in Wavelength zone, the upload bandwidth is quite high compared to the one to VM in availability zone. This shows that high-resolution video can be streamed to VM in Wavelength zone at high FPS, whereas it will be slow to stream to the VM in Availability zone. Download bandwidth from VM in Wavelength zone to device is quite high while from VM in Availability zone to device is relatively low. This indicates that it will not be possible to receive high network traffic from VM in availability zone to device, but it is possible to obtain from VM in Wavelength zone. If we see the upload and download bandwidth between VM in availability zone and VM in Wavelength zone, we can see that it is high enough to offload processing to availability zone VM and receive back responses on wavelength zone VM.

\rTable{network-performance-location-2} summarizes the latency, upload and download bandwidth measured between device in Wavelength zone to VM in Wavelength zone at location-2. We are not reporting (b) and (c) from \rTable{network-performance-location-1} in this table, since they are not directly affected by the 5G hotspot network performance at different locations. We notice that the latency is slightly lower, upload bandwidth is slightly higher, and download bandwidth is significantly higher at location-2 than location-1.

\subsection{Application-processing latencies}
\lSec{app-processing-latencies}
All measurements in \rSec{network-performance} were application-independent i.e. they were just related to network performance. In this section we report some application-specific performance numbers. Again, we run the application's microservices within two separate Kubernetes clusters in the availability and wavelength zone as shown in our test bed in \rFig{test-bed}. Our runtime manages deployment and co-ordination between these separate clusters.


\begin{figure}[t]
   \centering
   \includegraphics[width=0.95\linewidth]{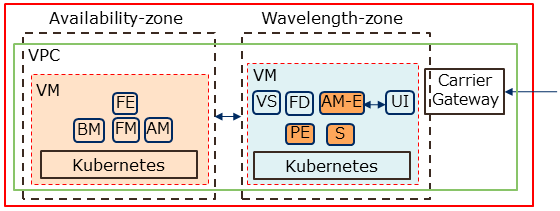}
   \caption{Hybrid deployment determined by our runtime}
   \lFig{hybrid}
\end{figure}

\begin{figure}[t]
\centering
    \includegraphics[width=0.95\linewidth]{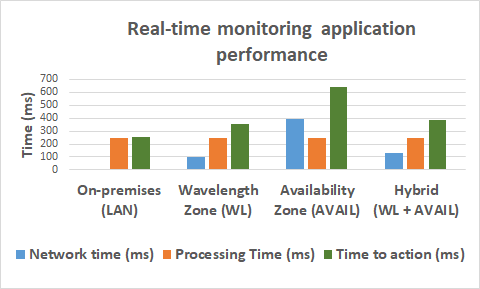}
    \caption{Real-time monitoring application performance}
    \lFig{real-time-monitoring-app-performance}
\end{figure}

\subsubsection{Real-time monitoring}
\rFig{real-time-monitoring-app-performance} shows the performance for real-time monitoring application for different deployment scenarios. Here we took one full HD frame and measured the time it takes to process this frame i.e. detect and match faces in the frame. This is the reported processing time. Note that we used similar hardware for different deployment scenarios and therefore processing time is the same across various scenarios. Next, we measured the network time i.e. the time spent in sending full HD frame over the network for processing and then receiving back the alert. We measured this time when processing was (a) On premises i.e. in local network (LAN) (b) On VM in Wavelength zone (edge-only) (c) On VM in Availability zone (cloud-only) and (d) Hybrid deployment (shown in \rFig{hybrid} determined by our runtime based on \rSec{policy-engine}). As expected, the total time to action for (a) is the least since everything is in the internal local network. Next is the time taken for (b), since Wavelength zone provides high bandwidth and low latency communication. This is followed by (d) which leverages the high bandwidth and low latency communication provided by Wavelength zone, and between Wavelength and Availability zone. This deployment reduces total cost (details provided in \rSec{cost}) by offloading some processing to availability zone at the expense of slightly increased total time to action, but still being within the latency constraint. Finally the time taken by (c) is the highest since the network performance for Availability zone is poor compared to Wavelength zone.

\subsubsection{Investigation and forensics}
\begin{table}[t]
\caption{Time (in seconds) to upload video files}
\begin{center}
    \begin{tabular}{|l|c|c|c|}
    \hline
     \makecell{Deployment \\ scenario} &  \makecell{Time to \\ upload \\ video-1 (s)} & \makecell{Time to \\ upload \\ video-2 (s)} & \makecell{Time to \\ upload \\ video-3 (s)} \\
    \hline
    (a) On Wavelength Zone & 553 & 33 & 61 \\
    (b) On Availability Zone & 3925 & 254 & 387 \\
    \hline
    \end{tabular}
\lTable{video-file-upload}
\end{center}
\end{table}

As mentioned in \rSec{investigation-forensics}, processing of archived video files for investigation and forensics purpose, depends on how fast the video files can be uploaded to the computing fabric, where actual processing would happen. \rTable{video-file-upload} shows the time it takes to upload three different video files to VM in Wavelength zone and VM in Availability zone. We see that the time to upload to Wavelength zone is an order of magnitude faster than uploading to the availability zone. As the upload continues, the processing starts immediately and intermediate results are received back to the devices in the carrier network. 
\rFig{investigation-processing} shows the overall processing of the video files in two different deployment scenarios i.e. on wavelength zone and availability zone. As shown in the figure, the uploading of file to wavelength zone is faster, compared to uploading on availability zone. By increasing the parallelism to adjust processing rate according to the upload rate, processing the video file is done much sooner (around 7x faster) on wavelength zone than on availability zone.

\begin{figure}[t]
\centering
    \includegraphics[width=0.95\linewidth]{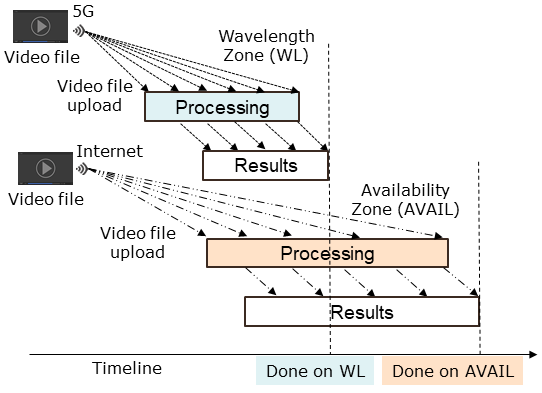}
    \caption{Investigation and forensics application processing on wavelength and availability zone}
    \lFig{investigation-processing}
\end{figure}


\subsection{Impact of runtime optimization}
\begin{figure*}[]
\begin{subfigure}[]{0.5\textwidth}
\centering
    \includegraphics[height=1.8 in]{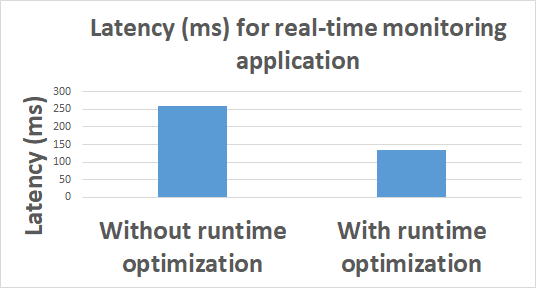}
    \caption{Static mapping}
    \lFig{runtime-static}
\end{subfigure}%
\begin{subfigure}[]{0.5\textwidth}
\centering
    \includegraphics[height=1.8 in]{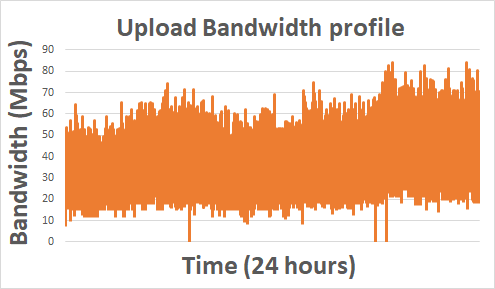}
    \caption{Bandwidth profile}
    \lFig{bandwidth-profile}
\end{subfigure}
\begin{subfigure}[]{0.5\textwidth}
\centering
    \includegraphics[height=2.0 in]{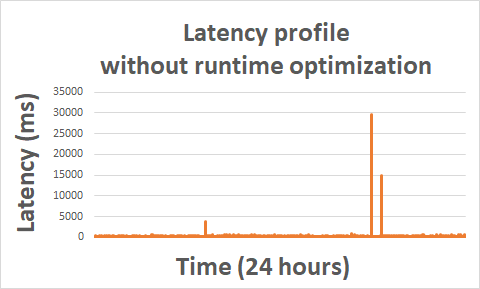}
    \caption{Dynamic mapping OFF}
    \lFig{dynamic-mapping-off}
\end{subfigure}%
\begin{subfigure}[]{0.5\textwidth}
\centering
    \includegraphics[height=1.85 in]{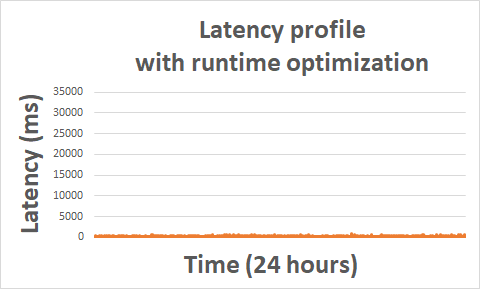}
    \caption{Dynamic mapping ON}
    \lFig{dynamic-mapping-on}
\end{subfigure}
\caption{Impact on latency}
\end{figure*}


In this section, we evaluate the impact of runtime optimization on application-processing latencies and cost. Particularly, we evaluate the scenario for static as well as dynamic mapping of microservices. We note that sometimes it is possible that our runtime cannot satisfy the response latency constraint, and in such cases this is notified to the developer. However, in our experiments such a scenario did not arise.

\subsubsection{Impact on static mapping}
\lSec{static}
\rFig{runtime-static} shows the network latency for static mapping of microservices of real-time monitoring application with and without runtime optimization. 
Without using our runtime, one way to deploy the application is to obtain the camera feed directly from carrier network into the Wavelength zone (leveraging low latency and high bandwidth offered by 5G) and then from there stream it into the cloud, where entire application processing happens (reducing cost of processing since VMs in cloud are cheaper) and then get back results through Wavelength zone VM back to the devices. For such a deployment, the overall network time is 260 milliseconds vs 134 milliseconds when hybrid deployment (see \rFig{hybrid}) determined by our runtime optimization is used. Our runtime optimization appropriately maps application's microservices on the edge and the cloud to keep the overall latency low.  

\subsubsection{Impact on dynamic mapping}
Along with the initial static mapping, as discussed in \rSec{scheduler} our runtime continuously monitors the application-level and network-level parameters and dynamically updates the mapping of microservices, if conditions change. To see this in action, we deployed the real-time monitoring application with latency constraint of 250 milliseconds, and  observed the application and network behavior over a period of 24 hours, with and without the dynamic mapping initiated by our runtime. As shown in \rFig{bandwidth-profile} we observe that the upload bandwidth between wavelength and availability zone drops to almost zero for 3 occasions in the 24 hour period. \rFig{dynamic-mapping-off} shows that for each of these three occasions, the latency suddenly spiked when dynamic mapping was turned OFF, whereas as shown in \rFig{dynamic-mapping-on}, when dynamic mapping was turned ON, the spike was handled seamlessly by our runtime by moving FE and FM microservices temporarily from availability zone to the wavelength zone, thereby keeping the end-to-end application latency below 250 milliseconds (specified constraint), despite network disruption.



\subsubsection{Impact on cost}
\lSec{cost}
Based on Amazon pricing \cite{amazon-pricing-link}, \rFig{cost-runtime} shows the impact our runtime optimization has on the cost of deployment of 100-cameras. For a static mapping as discussed in \rSec{static} without using our runtime, the number of VMs in Wavelength zone to stream the video to availability zone would be around 15 (required bitrate per camera to stream full HD video at 30 FPS is about 4 to 5 Mbits/s and total upload bandwidth between wavelength zone and availability zone is around 35 MBits/s, so 1 VM can support 7 cameras; to support 100 cameras around 15 VMs would be required) and number of VMs for processing in availability zone would be 50 (1 VM can support 2 cameras; so for 100 cameras 50 VMs would be needed), which brings the total cost per month to around \$9842.7, whereas the total cost per month for the hybrid deployment (see \rFig{hybrid}) determined by our runtime optimization would be around \$8253 per month. 
Our runtime optimization thus improves network latency by around 2x, while bringing the cost down by around 16\% for a 100-camera deployment.

\begin{figure}[t]
\centering
    \includegraphics[width=0.9\linewidth]{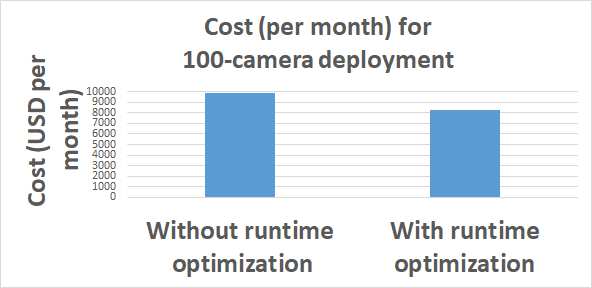}
    \caption{Impact on cost}
    \lFig{cost-runtime}
\end{figure}
\section{Related Work}
Microservices-based architecture has become quite popular in the past few years \cite{microservices-link} \cite{microservices} and is inspired by the concepts from service oriented architecture \cite{soa}. In this architectural style, monolithic applications are broken down into individual microservices, which are developed and deployed independently with lightweight interfaces between them for communication. This naturally results in loosely-coupled set of application microservices, which can be distributed across different computing fabric and can be scaled on-demand, making it a promising path towards implementing distributed IoT services \cite{microservices-iot}. Teemu et al.  \cite{edge-based-microservices-arch} showed such a path in designing and implementing distributed IoT applications using microservices. With typical IoT infrastructure, computing capability close to the source of data (devices) is fast but limited, while as you go further away e.g. into MEC \cite{MEC} and then into the cloud, the computing capability goes up, at the cost of higher latency \cite{edge-cloud-latency} \cite{edge-computing-overview}. 

With the flexibility provided by the microservices architecture, it is possible to map various components onto the device, MEC or the cloud. However, identifying which parts of the application needs to run where is a 
challenge \cite{edge-computing-challenges} and several partitioning and offloading techniques have been discussed by Jianyu Wang et al. \cite{edge-cloud-offloading-algorithms} and Varshney et al. \cite{Varshney_2020} in their survey paper. Techniques to optimally fuse and place functions in edge or cloud in serverless platforms have been proposed by Tarek Elgamel et al. \cite{costless}. Ren et al. \cite{collaborative-edge-cloud-computing} formulated the problem of splitting task between edge and cloud as purely latency minimization problem with constraints on communication and computation resources. Anirban Das et al. \cite{edge-cloud-perf-opt} present models to predict latencies and cost of running tasks in edge or cloud and use that to allocate task appropriately based on provided latency or cost constraints. However, they operate at a single serverless function level rather than looking at a pipeline of microservices to process an input. Techniques to offload some tasks of application from mobile devices to the cloud have been discussed in MAUI\cite{maui}, CloneCloud \cite{clonecloud}, and by Huaming Wu et al. \cite{partitioning-in-mobile-environment}. They aim to minimize energy consumption at mobile devices while keeping overall application latency low. They do not take into account the monetary cost of execution of tasks locally vs remotely. Moreover, the granularity at which tasks are offloaded is too fine e.g. at individual function-level. Game theory based approaches have also been proposed for optimizing deployments in edge-cloud infrastructure \cite{game-theory} \cite{yang2018noncooperative} \cite{ranadheera2017mobile}. Fard et al. \cite{moo} propose multi-objective scheduling of scientific workflows on heterogeneous computing environments, but their work focuses mainly on static scheduling and does not consider dynamic scheduling. 

With the advent of edge computing using 5G \cite{edge-computing-in-5g} \cite{aws-verizon-5g}, new applications are emerging and currently there is no standard way of programming and efficiently managing applications to take full advantage of this new vertically distributed, hierarchical, heterogeneous infrastructure with 5G connection at the edge. Maheswaran et al. \cite{language-edge-cloud} proposed a programming language with custom compiler to program applications for edge-cloud. This however, is quite restrictive in terms of deployment of generic microservices. Sumit et al. \cite{scalability-perf-eval-edge-cloud} propose model to provision resources in edge and cloud, but they consider connectivity between two edge-clouds, which does not exist in our setting, where 5G coverage is limited to certain area and only to certain Wavelength zones (direct low-latency connection between wavelength zones isn't currently available). Unlike any of existing work, we propose a runtime system, which enables applications to make effective use of 5G network, computing at the edge of 5G network, and computing in the cloud, while keeping the overall monetary cost of deployment and end-to-end application latency low.

\section{Conclusion}
Demand for edge computing is rising because centralized cloud computing with 100+ milliseconds network latencies  cannot meet the tens of milliseconds to sub-millisecond response times required for emerging 5G applications like autonomous driving, smart manufacturing, tactile internet, and augmented/virtual reality. 
In this paper, we present a new runtime that enables applications to make effective use of a 5G network, computing resources at the edge of this network,  and additional computing resources in the centralized cloud. Our dynamic, runtime optimizations leverages application-specific knowledge to improve application response latencies by 2x when compared with a static, manually optimized edge-cloud implementation on AWS Wavelength. Although we used carrier-supported 5G edge-cloud, our approach is equally applicable to local 5G implementations. As 5G networks continue to improve, and the demand for edge computing continues unabated, we expect the costs of edge-cloud to drop by an order of magnitude. This will usher in a new era where edge-cloud will be just as ubiquitous as cloud computing is today.

\bibliographystyle{IEEEtran}

\end{document}